# Talking Slide Avatars: Open-Source Multimodal Communication Approach for Teaching

Xinxing Wu [1]*

[1] School of Mathematics and Computer Science, Kentucky State University, Frankfort, Kentucky, United States

*Corresponding author: Xinxing Wu (xinxingwu@gmail.com; xinxing.wu@kysu.edu)

**Abstract**

Slide-based teaching is widely used in higher education, yet in online, hybrid, and asynchronous contexts, slides often lose instructor presence, narrative continuity, and expressive framing that help learners connect with course content. Full lecture video can partly restore these qualities, but it is time-consuming to record, revise, and reuse. This study presents a practice-based implementation and analytic reflection of an open-source workflow for creating talking slide avatars. The workflow integrates *OpenVoice* for text-to-speech and authorized voice-style conversion with *Ditto-TalkingHead* for audio-driven talking-image synthesis, enabling instructors to transform a short script and an authorized or synthetic portrait image into a narrated video for slide decks or HTML-based lecture materials. Rather than treating this workflow only as a technical solution, the study frames talking slide avatars as multimodal communication artifacts at the intersection of digital pedagogy, aesthetic education, and art-technology practice. The paper documents the production pipeline, analyzes communicative and aesthetic affordances, and proposes practical guidelines for script length, image selection, pacing, disclosure, accessibility, consent, and ethical use. Its contribution is not a validated learning intervention, but an educator-oriented open-source production model and communication-design framework. The study concludes that short, transparent, and carefully designed avatars may provide a reusable communication layer for introductions, transitions, reminders, and recaps when used selectively and with appropriate ethical safeguards.

## 1. Introduction

Slide-based teaching remains one of the most common approaches for organizing and presenting content in higher education,[1-2] yet the communicative experience of slides is often uneven across learning contexts. In face-to-face classrooms, an instructor's voice, pacing, emphasis, and body language can animate a slide deck, transforming static bullet points into a coherent and engaging explanation. In online, hybrid, and asynchronous settings, however, this live layer of instructional presence is frequently diminished, absent, or costly to reproduce.[3] Recorded lectures can partially mitigate this limitation, but full video production often demands repeated recording, editing, and revision whenever course content changes. For many instructors, particularly those managing multiple sections or updating materials each semester, this creates a practical tension between pedagogical goals and sustainable content production.[4] At the same time, research on multimedia learning and instructional video has shown that voice, temporal pacing, and social cues matter. Well-designed instructional videos can support engagement and comprehension, especially when they balance verbal explanation with clear visual focus.[5-8] Research on pedagogical agents further suggests that learners often interpret human-like on-screen agents socially rather than purely instrumentally, which means that voice, demeanor, and communicative style can shape how instructional messages are received.[9-12] Recent work on AI-based educational avatars extends this conversation by examining how synthetic speakers, embodied agents, and avatar-based interfaces may alter the experience of teaching and learning.[13-14]

This paper focuses on talking slide avatars for slide-based teaching, course lectures, and related forms of course communication. The workflow combines two open-source projects, *OpenVoice* and *Ditto-TalkingHead*, in an educator-friendly production process. Open-source components are used to generate speech from a script, optionally apply authorized voice-style conversion, animate a static portrait image, and export a short talking video that can be inserted into slides, course webpages, or learning management systems.[15-17] The central claim of this paper is that the significance of this workflow extends beyond its technical function. Talking slide avatars can be understood as communicative and aesthetic artifacts situated at the intersection of digital portraiture, speech performance, instructional design, and mediated presence. Their implementation also invites engagement with broader perspectives from art and technology, communication studies, and aesthetic education.

The goal of this paper is to offer an applied contribution to communication-oriented instructional design for education. The study is framed as a practice-based implementation and analytic reflection rather than as a controlled empirical evaluation of learning outcomes. Accordingly, the paper does not claim demonstrated improvements in learning, engagement, or usability; instead, it presents a reproducible workflow and a design framework that can inform future empirical evaluation. Specifically, the paper: (1) documents an open-source implementation pipeline; (2) analyzes the communicative, aesthetic, and pedagogical affordances of talking slide avatars; (3) proposes practical design guidelines for classroom use; and (4) addresses limitations, ethics, accessibility, consent, and future research directions. In doing so, the paper reframes talking avatars as a reusable layer of course communication that may support instructor presence and narrative continuity without requiring full-scale video production for entire lessons.

The remainder of the paper reviews the relevant scholarly context, documents the workflow, analyzes its prototype outcomes, considers broader implications for communication, ethics, and pedagogy, and concludes with limitations and future directions.

## 2. Conceptual and scholarly context

Recent scholarship on generative AI in education emphasizes that educational value depends not only on technical capability, but also on teacher adoption, responsible use, transparency, accessibility, and pedagogical alignment. This perspective is important for the present study because talking slide avatars should not be evaluated merely as technical outputs, but as designed communication artifacts situated within changing practices of digital teaching and mediated presence.

This paper draws on three overlapping areas: multimedia learning, pedagogical agents, and contemporary AI-avatar research. First, multimedia learning research has long emphasized that students learn from coordinated combinations of words and images rather than from either channel alone. Instructional videos extend this principle by adding timing, narration, pacing, and attention guidance. Studies of educational video production have shown that relatively small design choices, such as clip length, segmentation, and conversational delivery, can have measurable implications for engagement. In addition, systematic reviews of instructor presence in instructional videos show a more nuanced picture: visually present instructors do not automatically improve learning outcomes, yet carefully designed social and attentional cues can increase affective engagement and support aspects of learning in specific conditions.

Second, research on pedagogical agents suggests that learners often respond to computer-based agents as social actors. Animated or embodied instructional agents may trigger what has been described as social agency, in which learners process the message as if they are in a form of guided interaction rather than simply receiving neutral information. This does not mean that any animated face is educationally beneficial; indeed, agent design effects are often mixed and highly dependent on appearance, voice, context, and task type. What the literature does support, however, is the idea that voice quality, politeness, social cues, and a coherent persona matter in multimedia environments. For a project such as talking slide avatars, this means the design question is not "*Can a face move with speech?*" but "*What kind of communicative relationship does the avatar establish with the learner?*"

Third, the rapid development of generative AI has made synthetic avatar production more accessible. Recent scholarship has begun to examine AI-based avatars as educational tools, highlighting both their potential and their risks. More broadly, work on generative AI in education emphasizes the opportunities and practical concerns associated with integrating AI tools into teaching, including transparency, responsible adoption, and pedagogical alignment. Although Uğraş et al.[32] focus on ChatGPT rather than avatar-based media, their study reflects the broader educational shift toward AI-supported instructional communication. One empirical study comparing video lectures with human instructors and AI-generated instructors found lower engagement for the AI-generated version but comparable learning performance, suggesting that technical realism and social acceptance still shape reception even when content equivalence is preserved.[18] This finding pushes design beyond novelty. The challenge is not simply to create an avatar, but to create one whose visual and vocal behavior supports trust, clarity, and communicative purpose.

Within the broader field of arts and communication, AI-generated media are increasingly recognized as transformative forces in image-making, creativity, and cultural production. Recent scholarship on AI, digital art, and mediated communication suggests that generative systems are not simply instruments of automation; they also reconfigure aesthetic authorship, mediation, and reception.[19–23] The talking slide avatar belongs to this broader landscape. It is neither a conventional lecture recording nor merely a software output. Rather, it is a designed performance artifact: a synthetic micro-presentation in which voice, face, timing, and slide content are intentionally orchestrated to communicate meaning. This artistic-communicative framing is particularly relevant in educational settings, where the design of explanation is inseparable from the design of attention, engagement, and trust.

The present study contributes to this conversation by offering a concrete, reusable, and comparatively low-cost open-source approach to producing such artifacts. Unlike proprietary avatar-generation platforms that obscure production logic behind closed interfaces, the workflow developed here renders the process visible and editable through scripts, notebooks, and documented procedures. Such transparency matters both academically and pedagogically, because it allows the artifact to be analyzed, adapted, critiqued, and repurposed not only as a teaching resource, but also as a case of communication-oriented design.

Compared with empirical studies of AI-generated instructors and pedagogical avatars, the present study has a different scope. Prior empirical work has examined how learners respond to AI-generated instructors, humanoid avatars, or instructor presence in video lectures, often using engagement, performance, or perception measures. By contrast, the present study does not test learner outcomes. Its contribution is an open-source implementation and design analysis of a practical workflow for producing short talking-avatar clips that can be embedded in slide-based learning materials. This distinction positions the workflow as a reproducible communication-design resource rather than as a validated instructional intervention.

## 3. Materials and methods

This study adopts a practice-based implementation and analytic reflection approach. Its purpose is to document, interpret, and critically analyze an open-source workflow for producing talking slide avatars as communication artifacts for teaching. The analysis focuses on workflow feasibility, reproducibility, communicative affordances, design implications, and ethical safeguards. No survey data, classroom experiments, controlled comparisons, usability testing, or human-subject outcome measures are reported.

The implementation accompanying this study includes a README, a text-to-speech notebook (*TextToVoice.ipynb*), a talking-image notebook (*ImageSpeaking.ipynb*), supporting dependencies, and integrated open-source components related to *OpenVoice* and *Ditto-TalkingHead*. The purpose is not to analyze the repository as an object in itself, but to document how the workflow translates a written lecture script into a speaking slide artifact and to examine the communicative design possibilities that such a workflow makes available.

As shown in Figure 1, the workflow consists of four stages. First, the instructor prepares a short script for a slide segment, such as a weekly introduction, an assignment explanation, or a concept summary. Second, the script is converted to speech using *OpenVoice*. In the implementation provided with this paper, the workflow initializes the BaseSpeakerTTS model and the ToneColorConverter, selects an English base speaker checkpoint, and, when desired, extracts speaker embeddings from a reference audio clip to approximate a preferred vocal style. The text is then rendered into a temporary waveform and converted into a cloned or style-adjusted output audio file. Third, the generated audio is combined with a portrait image in *Ditto-TalkingHead*. In the implementation, this stage uses a *Google Colab*-based setup, downloads model checkpoints, installs dependencies, and runs an inference process to synthesize an MP4 talking-avatar clip from the audio-image pair. Fourth, the resulting short video is embedded into lecture slides or HTML-based course pages, where it functions as a compact speaking element rather than as a full-length lecture recording.

Several implementation characteristics are especially important. The notebooks are structured for *Google Colab* and *Google Drive*, which reduces local installation demands for instructors who may not wish to maintain a dedicated machine-learning environment. The workflow also separates speech generation from facial animation, making the process modular. This modularity is pedagogically useful because instructors can revise scripts, change a voice reference, or replace the avatar image without redesigning the entire presentation. In addition, the implementation recommends short clips, clear front-facing images, and moderate speech pace, all of which align with established guidance on instructional video segmentation and attention management.

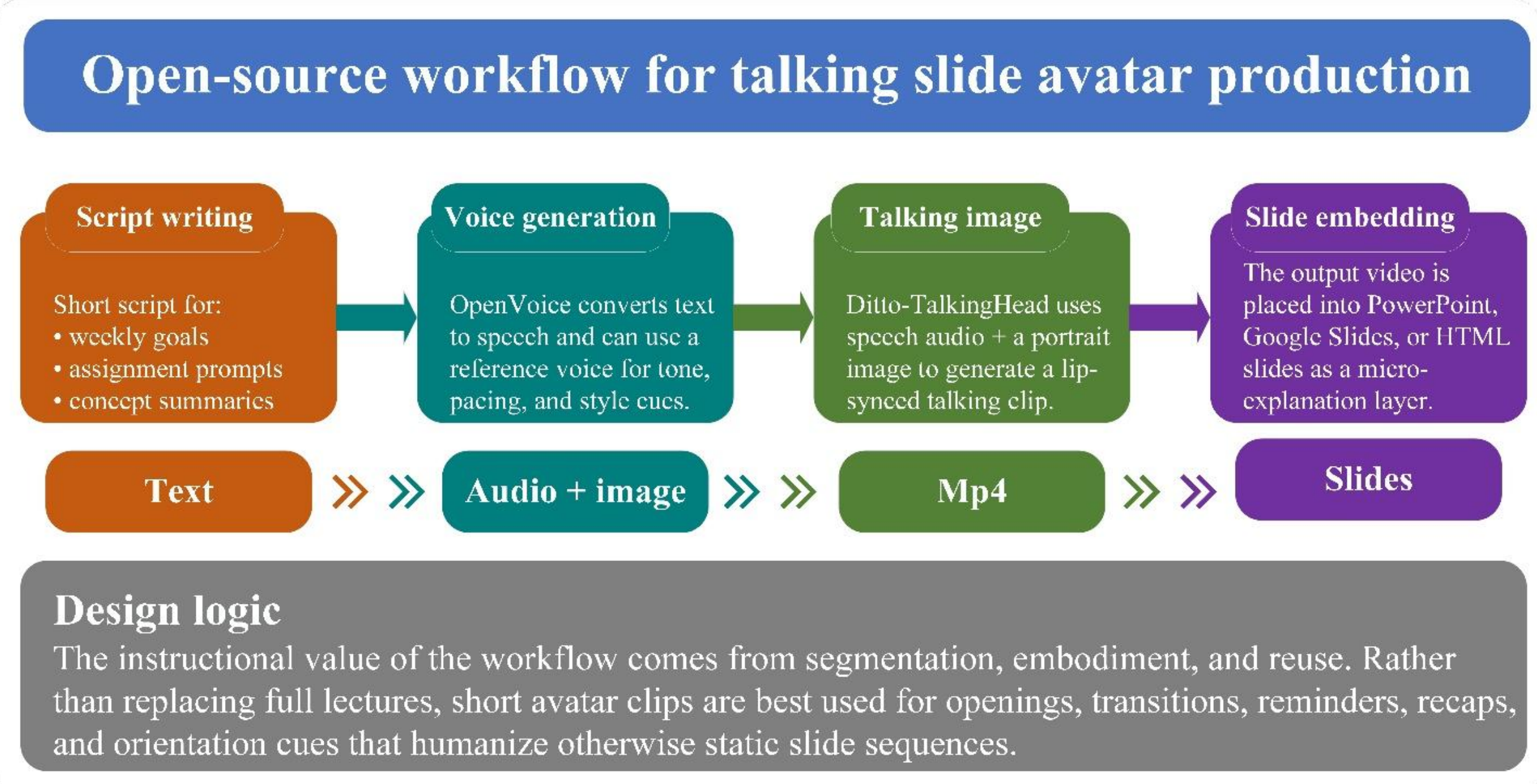

**Figure 1. Workflow diagram for talking slide avatar production. The figure was created to summarize the open-source workflow described in this study.**

The analytic reflection in this paper is organized around three dimensions. The first is technical reproducibility: whether the workflow is documented clearly enough to be followed, adapted, and reused by educators. The second is communicative function: what kinds of instructional messages the talking-avatar format is particularly well suited to deliver. The third is aesthetic-pedagogical design: how the synthetic combination of image, speech, timing, and slide context shapes the tone, presence, and meaning of the communication. Because this paper does not report human-subject data, survey data, or classroom experiments, it does not claim measured learning gains. Instead, it offers an original implementation account together with a theoretically grounded discussion of why this workflow matters for educational communication.

The technical implementation is intentionally concrete. In the text-to-speech stage, the workflow mounts cloud storage, clones the *OpenVoice* repository, installs required dependencies, loads checkpoints for the English base speaker and converter, generates a temporary waveform from the instructional script, and then applies tone-color conversion using a reference speaker clip. In the talking-image stage, the workflow mounts storage again, clones *Ditto-TalkingHead*, downloads checkpoints, installs PyTorch and supporting packages, and runs an offline inference command that takes a source image, an audio waveform, and an output path as its core inputs. The separation of these stages is more than a coding convenience; it creates a pipeline that can be documented, taught, maintained, and revised in modular form.

Table 1 summarizes the key components of the workflow and their roles in the implementation. Table 2 provides additional reproducibility information, including the public repository, demo page, notebooks, runtime environment, required inputs, outputs, and ethical documentation.

**Table 1. Summary of implementation components.**

| Component | Role in workflow | Implementation in This Study |
|---|---|---|
| Script | Defines the message to be spoken | Short lecture text prepared by the instructor for introductions, assignment explanations, or concept summaries. |
| *OpenVoice* | Text-to-speech generation and voice-style conversion | BaseSpeakerTTS generates source audio; ToneColorConverter and speaker embedding extraction allow style transfer from a reference clip. |
| Reference audio | Shapes the vocal color or identity of the output | An authorized reference audio file may be used to extract speaker embeddings for consented voice-style conversion. |
| *Ditto-TalkingHead* | Audio-driven talking image synthesis | The inference pipeline combines a source image and generated WAV audio to produce an MP4 talking-avatar clip. |
| Portrait image | Provides the visual anchor for embodiment | A clear, front-facing, synthetic/non-identifiable, author-owned, openly licensed, public-domain with documented provenance, or consented image is recommended to improve lip-sync plausibility, visual stability, and responsible use. |
| Slide or HTML page | Presentation context for the avatar | The final MP4 can be embedded into PowerPoint, Google Slides, or HTML-based course slides. |

**Table 2. Reproducibility summary for the open-source workflow.**

| Reproducibility item | Documentation in this study |
|---|---|
| Public repository | VirtualAssistant repository: https://github.com/xinxingwu-uk/VirtualAssistant |
| Public demo | Updated demo: https://xinxingwu-uk.github.io/projects/demo3_updated/slides.html |
| Main notebooks | TextToVoice.ipynb for speech generation; ImageSpeaking.ipynb for talking-image synthesis |
| Core components | OpenVoice for text-to-speech and voice-style conversion; *Ditto-TalkingHead* for audio-driven talking-image generation |
| Runtime environment | *Google Colab* and *Google Drive* workflow |
| Required inputs | Instructional script, authorized/reference voice audio if voice-style conversion is used, and an authorized or synthetic/non-identifiable avatar image |
| Main outputs | WAV audio file and MP4 talking-avatar clip |
| Reuse pathway | Generated MP4 can be embedded into PowerPoint, Google Slides, or HTML-based course slides |
| Ethical documentation | Image source/provenance, voice consent, AI-generation disclosure, captions/transcripts where possible |

## 4. Prototype outcomes and design analysis

This section reports prototype outcomes and design observations rather than measured learning or usability results. The primary outcome of the implementation is a repeatable production model for creating short talking slide avatars. Many educators do not need a fully autonomous conversational avatar; rather, they need a lightweight and sustainable method for adding brief spoken presence to otherwise static slide sequences. The workflow addresses this instructional niche by supporting reusable avatar clips for openings, transitions, reminders, and recaps.

### 4.1 Instructional use cases

From a communication-design perspective, the workflow appears particularly well suited to four scenarios. The first is slide opening and orientation. A short avatar clip can welcome students, introduce the agenda, or frame expectations before the main content begins. The second is transition management. In longer slide decks, the avatar can signal movement from one conceptual block to another, reducing the abruptness that often characterizes asynchronous slide viewing. The third is task explanation. Assignment instructions, weekly reminders, and logistical updates are often too brief to justify full video production, but they may benefit from concise spoken framing. The fourth is emphasis and recap. A short speaking avatar at the end of a section may help reinforce a key takeaway and provide a more personable closing cue than a final static bullet list alone.

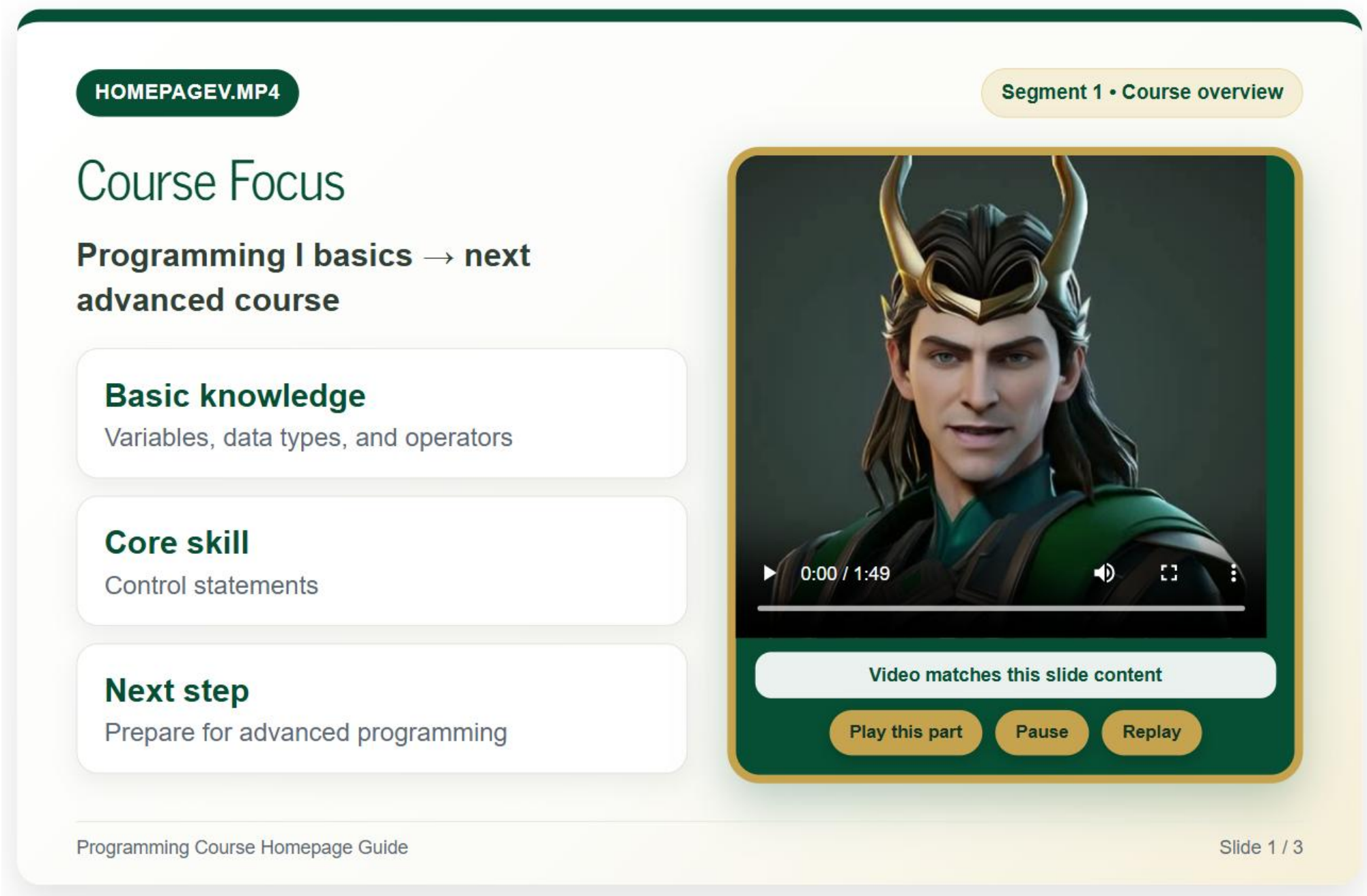

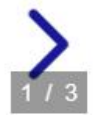

**Figure 2. Screenshot of a talking slide avatar embedded in an HTML-based lecture slide. The figure was created by the author from the updated public demo and uses a synthetic, non-identifiable fantasy-style avatar for demonstration purposes. The avatar functions as a compact communication layer for orientation, transition, reminder, and recap purposes rather than as a replacement for full lecture video.**

To avoid ambiguity regarding portrait-image provenance and likeness rights, the revised prototype uses a synthetic, non-identifiable fantasy-style avatar rather than a recognizable historical or public figure. The example is intended to demonstrate the technical and communicative structure of the workflow, not to recommend the use of public figures or identifiable persons in routine educational deployment.

The implementation also highlights an important artistic dimension. The avatar clip functions less as a neutral narration layer than as a micro-performance. The generated voice is timed, stylized, and embodied through a portrait image. When embedded within slides, the clip transforms a conventional presentation into a mixed-media object composed of text, image, recorded sound, and synthetic animation. In this sense, the instructor is not only a content author, but also a designer of mediated presence.

At the same time, the implementation reveals several design boundaries. The first is duration. The accompanying workflow recommends short clips of approximately 10 to 30 seconds, and this recommendation is well justified. Short duration helps preserve novelty, reduces rendering cost, minimizes the visibility of animation artifacts, and aligns with established guidance favoring segmented instructional video over long, uninterrupted presentation streams. The second boundary is image quality. The avatar performs most effectively when driven by a clear, front-facing image with stable lighting and minimal occlusion. This requirement is both technical and aesthetic:

weaker source images not only reduce lip-sync plausibility, but may also diminish perceived trustworthiness and visual harmony with the surrounding slide design. A third boundary concerns realism. Prior work on AI-generated instructors suggests that imperfect facial animation can lower engagement even when instructional performance remains broadly comparable. That finding helps explain why the talking slide avatar is best positioned as a supplementary communicative layer rather than a complete substitute for live teaching or full recorded lecture presence. The format appears most appropriate when assigned specific, bounded functions such as orientation, transition, emphasis, or recap. It is less convincing when expected to sustain extended explanation in a highly human-realistic style.

Taken together, these observations suggest that the potential value of the prototype lies not in replacing teaching, but in offering a scalable and reusable layer of voice, face, and timing for slide-based communication. Its strongest contribution is therefore design-oriented: it offers a practical model for introducing mediated instructor presence into slide environments in ways that are lightweight, adaptable, and aesthetically coherent.

### 4.2 Modularity, continuity, and rhetorical design

The workflow also changes how slide communication is designed. Because the script, voice, portrait image, and embedding context remain separable, instructors can revise one component without rebuilding the entire artifact. This supports both practical reuse and aesthetic continuity. For example, an instructor may keep a consistent avatar image and voice style across a course while changing the weekly script, or may vary the avatar design across modules while preserving a recognizable format.

The format also encourages a rhetorical shift from dense slide prose to short, speakable language. Traditional slide text is often compressed for silent reading, whereas avatar scripts must be paced naturally and understood through listening. This shift may improve clarity and tone, although its effect on learner experience requires future empirical study. Table 3 summarizes design guidelines for classroom deployment.

**Table 3. Design guidelines for classroom deployment.**

| Design issue | Recommendation | Rationale |
|---|---|---|
| Clip length | Keep most clips within 10-30 seconds. | Short clips preserve attention, reduce rendering time, and align with segmented instructional video design. |
| Message type | Use avatars for openings, transitions, reminders, and recaps rather than dense full lectures. | The format is strongest for bounded communication tasks, not long continuous exposition. |
| Image selection | Use front-facing images with clear lighting and minimal occlusion. | Better source images improve both technical performance and perceived trust. |
| Voice design | Prefer natural pacing and moderate speech speed; avoid overly dramatic tone shifts. | A stable and conversational voice reduces distraction and supports message clarity. |

| Design issue | Recommendation | Rationale |
| --- | --- | --- |
| Disclosure | Label clips as AI-assisted or AI-generated where appropriate. | Transparency helps protect trust and distinguishes legitimate educational media from deceptive synthetic content. |
| Accessibility | Provide captions or transcripts whenever possible. | Students benefit from multimodal access and review options. |
| Aesthetic consistency | Reuse a small set of coherent avatar styles across a course. | Consistency helps the avatar function as a recognizable communicative motif rather than a novelty. |
| Frequency of use | Insert selectively at key slide moments. | Overuse may create redundancy, fatigue, or unnecessary cognitive load. |
| Consent and identity | Use only authorized, consented, openly licensed, public-domain with documented provenance, author-owned, or synthetic/non-identifiable images and voices. | Ethical deployment requires clear control over likeness, voice identity, authorship, privacy, licensing, disclosure, and permission. |

## 5. Broader implications for communication, ethics, and pedagogy

The workflow has broader implications for communication, ethics, and pedagogy because it treats instructional materials as hybrid media artifacts. A talking slide avatar is shaped through the coordinated design of script, voice, portrait image, timing, animation, and slide context. It is therefore not merely a software output, but a compact synthetic performance situated at the intersection of education, media design, and synthetic representation. This framing connects the workflow to wider discussions of authorship, mediation, trust, and aesthetic production in AI-generated media. [24-30]

### 5.1 Openness, sustainability, and pedagogical reflexivity

Many commercial avatar platforms prioritize convenience but conceal the production process behind closed interfaces. By contrast, the workflow presented here remains comparatively transparent: scripts can be revised, voice references can be inspected, dependencies are explicit, and outputs are generated through documented steps. This openness supports adaptation, critique, and reuse by educators and researchers.

The workflow may also support teaching sustainability. Short avatar clips can reduce the need to repeatedly record routine course messages, such as weekly introductions, assignment reminders, or brief transitions. At the same time, responsible use requires disclosure, permission practices for voice and image sources, accessibility support, and continued instructor accountability for the message being communicated.

### 5.2 Trust, disclosure, and realism

Trust is especially important because synthetic instructional media can approach the communicative territory of deepfakes and other forms of deceptive representation. In educational settings, the distinction between acceptable synthetic assistance and problematic deception depends not only on the model itself, but also on context, disclosure, authorship, and intent. A transparently labeled avatar used by an instructor to present a scripted course message is ethically distinct from an unlabeled synthetic representation that implies a naturally recorded or live performance. The workflow is therefore strongest when it presents the avatar as a clearly designed media object rather than as a hidden simulation of unmediated presence.

Questions of realism introduce a further communicative and ethical consideration. Research on avatar realism and humanoid interfaces suggests that highly human-like agents may become less effective when facial expression, timing, or motion appear almost realistic but not fully convincing, thereby producing distraction or unease.[23,29] For course communication, the design goal should not necessarily be maximal realism. In many situations, a modestly stylized, clearly artificial, and visually coherent avatar may be more appropriate than one that strives for perfect human substitution. This reinforces the argument that talking slide avatars should be understood as communication artifacts whose usefulness depends on aesthetic fit, rhetorical function, transparency, and contextual appropriateness rather than on model fidelity alone.

### 5.3 Image provenance, voice consent, and responsible use

Although the workflow described in this study uses open-source software components, open-source availability of code does not remove the need to respect privacy, copyright, publicity rights, and consent. The workflow does not require student images, student voices, student records, or identifiable student data. For responsible use, portrait images should be author-owned, openly licensed, public-domain with documented provenance, consented, or synthetic/non-identifiable demonstration images. Recognizable likenesses of living persons, students, instructors, public figures, historical figures, or protected fictional characters should not be used without permission or a clearly documented legal basis.

Voice cloning requires similar caution. Reference audio should be limited to the instructor's own voice, openly licensed voice samples, or audio for which explicit consent has been obtained. The workflow should not be used to imitate another person's voice without authorization or in any way that could mislead students, colleagues, or the public. In educational deployment, AI-assisted or AI-generated avatar clips should be transparently labeled, and captions or transcripts should be provided whenever possible. These safeguards help distinguish legitimate educational synthetic media from deceptive or unauthorized representations.

### 5.4 Authorship and broader applications

From a pedagogical standpoint, the workflow is most appropriate when used selectively. Talking slide avatars are better suited for short openings, transitions, reminders, and recaps than for replacing full lectures or sustained instructor explanation. This use aligns with multimedia learning and instructional video research, which suggests that segmentation, signaling, pacing, and purposeful social cues are generally more effective than simply increasing media density.[30-31]

The workflow also raises questions of authorship. In conventional slide-based teaching, the instructor usually serves as the content author, while the software functions mainly as a delivery medium. In a talking-avatar workflow, however, the final artifact emerges from the combined action of instructor decisions, scripted language, image selection, voice-reference choices, and model behavior. This does not erase human authorship, but it makes authorship more distributed and mediated. For scholarship in art, media, and communication, this hybrid authorship is relevant because generative systems increasingly reshape creative practice, mediation, and aesthetic reception.[24-30]

Beyond conventional lectures, the workflow could be adapted for museum micro-explanations, student project showcases, digital exhibitions, conference posters with embedded narration, or multilingual public-facing educational materials. Because the artifact is essentially a short narrated speaking image, it can travel across educational and cultural communication settings with relatively little modification. Its broader value therefore lies not in replacing teachers or live communication, but in offering a reusable and ethically governed format for compact synthetic-media communication.

## 6. Limitations and future directions

This study has several limitations. First, it is an implementation-and-design analysis rather than a controlled classroom experiment, and it therefore does not provide direct evidence of student learning outcomes, satisfaction, engagement, retention, or usability. Second, the discussion is based on one implementation configuration centered on *OpenVoice* and *Ditto-TalkingHead*; future tools may alter the balance among output quality, production speed, controllability, accessibility, and ease of use. Third, the paper focuses on short, portrait-based avatar clips and does not examine conversational agents, interactive avatar systems, or fully dynamic video lecturers. Finally, because the workflow is demonstrated through a prototype rather than through classroom deployment, its practical value should be interpreted as a design-oriented contribution rather than as evidence of pedagogical effectiveness.

Future research should evaluate the workflow through a staged design. First, structured expert review could ask instructional designers, accessibility specialists, communication scholars, and AI ethics reviewers to assess usability, clarity, disclosure, accessibility, and ethical safeguards. Second, small pilot studies could collect learner feedback on perceived clarity, instructor presence, trust, cognitive load, and usability. Third, controlled comparisons could examine static slides, audio-only narration, human-recorded video, and AI-generated talking slide avatars using measures of engagement, comprehension, retention, and learner preference. These studies would allow the workflow to move from a practice-based implementation toward an evidence-supported pedagogical framework.

Additional future work should strengthen accessibility through captions, transcripts, language switching, and playback controls. Governance will also become increasingly important as avatar production becomes more accessible, requiring institutional policies for consent, disclosure, archival transparency, and acceptable use in formal teaching materials.

## 7. Conclusion

This study presented an open-source workflow for producing talking slide avatars through script writing, synthetic speech generation, talking-image synthesis, and slide integration. The workflow occupies a practical middle ground between static slides and full lecture video by enabling instructors to create short, reusable avatar clips for introductions, transitions, reminders, and recaps.

The contribution of the paper is not a new algorithm or a validated learning intervention, but a reproducible implementation and communication-design framework. By framing talking slide avatars as multimodal communication artifacts, the study highlights how timing, tone, image, voice, disclosure, accessibility, and ethical safeguards shape the educational meaning of synthetic media.

Talking slide avatars should not be understood as replacements for teachers or as automatically effective instructional tools. Their value depends on selective use, transparent labeling, consented or synthetic media sources, accessibility support, and alignment with specific communication purposes. Future empirical and expert-evaluation studies should examine how learners and instructors experience these artifacts and whether they improve engagement, usability, comprehension, or retention. With this scope in mind, the present study offers a foundation for responsible, reproducible, and ethically governed synthetic media in educational communication.

## Availability of data

The implementation accompanying this study is publicly available in the VirtualAssistant repository: https://github.com/xinxingwu-uk/VirtualAssistant . A public demo is also available at https://xinxingwu-uk.github.io/projects/demo3_updated/slides.html The repository includes the main workflow materials described in the paper, including notebooks for speech generation and talking-image synthesis.